\documentclass[10pt,twocolumn,conference]{IEEEtran}

\usepackage{amsmath}
\usepackage{amssymb}
\usepackage[dvips]{graphicx}
\usepackage{setspace}
\usepackage{epsfig}
\usepackage{amsmath}
\usepackage{amssymb}
\usepackage[dvips]{graphicx}
\usepackage{epsfig}
\usepackage{color}
\usepackage{cite}

\newcommand{\tSNR}{\text{SNR}}

\newtheorem{theo}{Theorem}

\newtheorem{lemm}{Lemma}
\newcommand{\figsize}{0.38}

\newcommand{\ssp}{\text{sp}}
\newcommand{\diag}{\text{diag}}

\definecolor{blue}{rgb}{0,0,1}

\begin{document}
\IEEEoverridecommandlockouts
\title{Backlog and Delay Reasoning in HARQ Systems}
\author{\authorblockN{Sami Akin and Markus Fidler}
\authorblockA{Institute of Communications Technology\\
Leibniz Universit\"{a}t Hannover\\
Email: \{sami.akin and markus.fidler\}@ikt.uni-hannover.de}
\thanks{This work was supported by the European Research Council under Starting Grant--306644.}}
\date{}

\maketitle

\begin{abstract}
Recently, hybrid-automatic-repeat-request (HARQ) systems have been favored in particular state-of-the-art communications systems since they provide the practicality of error detections and corrections aligned with repeat-requests when needed at receivers. The queueing characteristics of these systems have taken considerable focus since the current technology demands data transmissions with a minimum delay provisioning. In this paper, we investigate the effects of physical layer characteristics on data link layer performance in a general class of HARQ systems. Constructing a state transition model that combines queue activity at a transmitter and decoding efficiency at a receiver, we identify the probability of clearing the queue at the transmitter and the packet-loss probability at the receiver. We determine the effective capacity that yields the maximum feasible data arrival rate at the queue under quality-of-service constraints. In addition, we put forward non-asymptotic backlog and delay bounds. Finally, regarding three different HARQ protocols, namely Type-I HARQ, HARQ-chase combining (HARQ-CC) and HARQ-incremental redundancy (HARQ-IR), we show the superiority of HARQ-IR in delay robustness over the others. However, we further observe that the performance gap between HARQ-CC and HARQ-IR is quite negligible in certain cases. The novelty of our paper is a general cross-layer analysis of these systems, considering encoding/decoding in the physical layer and delay aspects in the data-link layer.
\end{abstract}

\section{Introduction}
Due to the dynamic nature of wireless media, transmission errors easily occur, especially when channel fading is too strong. Hence, in order to ensure an error-free (or error-minimized) transmission, error detection and correction techniques such as automatic-repeat-request (ARQ) and forward-error-correction (FEC) are engaged in certain classes of communications systems. Specifically in ARQ-furnished systems, the receiver feeds the transmitter with either a positive acknowledgment (ACK) or a negative ACK (NACK) after each data packet is received correctly or incorrectly so that the transmitter can either send the next packet or resend the same packet, respectively. On the other hand, in FEC each packet is encoded by introducing redundant data bits for the receiver to detect and correct errors. While FEC provides a fixed transmission rate, it does not adjust error correction capacity with respect to channel variations. However, ARQ can adapt to the channel variations by resending the transmitted packets. In that respect in ARQ systems, the transmission of a packet can be completed in a shorter period if the channel attenuation is weak, whereas it can be completed in a longer period if the channel attenuation is strong. Thus, in order to provide better error correction performance, FEC and ARQ have been merged to form hybrid-ARQ (HARQ).

Several HARQ protocols have been suggested and analyzed. For instance, originally proposed by Wozencraft and Horstein \cite{horstein_1}, Type-I HARQ (HARQ-T1) relies on a blended process of error detections and error corrections while the retransmission of one packet is initiated if this process fails. Subsequently, unlike HARQ-T1 in which the past erroneously acquired signals are discarded, Type-II HARQ was introduced in \cite{lin}. Additional information is transmitted to the receiver when the receiver fails decoding the acquired signal. On that account, one of the leading studies introduced an adaptive-feedback coding strategy that performs Type-II HARQ with incremental redundancy (HARQ-IR) \cite{mandelbaum}. Likewise, a widely-known method of packet combining technique, named Type-II HARQ chase combining (HARQ-CC), has been analyzed in \cite{chase}. In HARQ-CC, the repeated transmissions of a packet, before being decoded, are linearly added in order to obtain a power gain. From then on, Type-I and Type-II HARQ protocols have been analyzed from various perspectives such as coding performance \cite{f_cheng,shea,malkam}, outage probability \cite{wu_jin}, and power optimization \cite{djonin,chaitanya}. Furthermore, a link-level performance comparison of these two techniques has been investigated in \cite{frenger}, and it was shown that HARQ-IR can be significantly better than HARQ-CC, especially for high channel-coding rates and high modulation orders.

Meanwhile, since HARQ systems may require the retransmissions of packets, delay-sensitive traffic concerns have become a research focus as well. For instance, considering adaptive modulation techniques, Villa \emph{et al.} investigated latency-constrained networks \cite{tania}. In another line of research, the author in \cite{anasto} formulated the design of an HARQ system as a stochastic control problem where the transmitter controls the code rate of each packet in order to achieve a minimum overall average delay. Similarly in \cite{gunaseelan}, effective capacity \cite{wu_negi}, a dual of effective bandwidth \cite{c_chang}, has been characterized as the performance metric. A two-dimensional continuous-time Markov channel model has been employed in order to partition the instantaneous data rate received at the destination into a finite number of states each of which is representing a mode of operation of the HARQ scheme. In addition, Huang \emph{et al.} conducted the analysis of queueing performance of HARQ systems by modeling the wireless channel as a two-state continuous Markov process and derived the queueing delay from generating functions of the finite-buffer-capacity system \cite{huang}. Using the framework of the statistical network calculus, the authors investigated the delay performance of block codes with more or less redundancy \cite{lubben}. Finally, we note that taking the first order expansion of the effective capacity under loose quality-of-service (QoS) constraints into account, the authors in \cite{ozcan} studied the impact of deadline constraints, outage probability, and QoS constraints on the energy efficiency in HARQ systems, and compared HARQ-CC and HARQ-IR protocols.

\begin{figure}
\begin{center}
\includegraphics[width=0.35\textwidth]{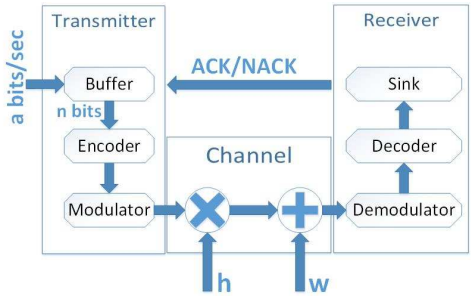}
\caption{Channel model. The transmitter initially stores the data packets in its buffer, subsequently encodes, modulates and forwards each data packet to the receiver through the wireless channel. Then, the receiver feeds the transmitter with ACK/NACK depending on its decoding performance.}\label{res:res_1}
\end{center}
\end{figure}
Different than the above studies, we evaluate the queueing performance of a broad range of HARQ wireless systems by modeling the transmitter queue activity aligned with the decoding performance at the receiver as a discrete $M$-state Markov process with a transmission deadline strategy for each data packet. To the best of our knowledge, a cross-layer examination of a general class of HARQ systems, regarding the effect of wireless channel attenuation on decoding performance in physical layer together with queueing performance in data-link layer, has not been performed. Singly, we can express our contributions with the following:
\begin{itemize}
  \item We identify an approach and construct a state-transition model for any HARQ system to investigate how advanced transmission schemes at the physical layer affect the queueing performance at the data link layer.
  \item Using the above model, we analyze the interplay between the packet-loss probability at the receiver due to transmission deadlines, and the steady-state probability of clearing the transmitter queue.
  \item We obtain a closed-form expression for the effective capacity of HARQ systems, which provides the maximum feasible arrival rate at the transmitter queue under QoS constraints such as the asymptotic probabilities of buffer overflow and delay violation.
  \item Considering the stability condition of the queue, we analyze the non-asymptotic backlog and delay bounds of these systems.
\end{itemize}

In this paper, we provide a general framework to investigate the interactions between the QoS concerns, and the coding and modulation techniques embedded in HARQ protocols. For instance, the adoption of linear coding and modulation schemes in HARQ protocols and their delay performance can be easily analyzed using this conduct. The organization of the rest of the paper is as follows. In Section \ref{system_model}, we describe the system model. Then, we discuss in Section \ref{system_throughput} the system performance regarding packet-loss probability, effective capacity, and non-asymptotic performance bounds. Finally, we provide the conclusion in Section \ref{conclusion}, and the relegated proofs in Appendices \ref{app:effective_capacity} and \ref{app:slack_term}.

\section{System Model}\label{system_model}
In this section, we present in detail the transmission and channel model, a general view of HARQ systems, and the state transition model of these systems. 
\subsection{Transmission and Channel Model}\label{transmission_model}
We consider a point-to-point transmission system in which one transmitter and one receiver communicate over a wireless fading channel as seen in Figure \ref{res:res_1}. In this model, it is assumed that the transmitter initially divides the available data into packets of $n$ bits and stores them in its buffer. Then, it performs the encoding, modulation, and transmission of each packet in time-slots of $T$ seconds with the first-come first-served policy. It is further assumed that each packet should be transmitted in maximum $MT$ seconds due to a transmission deadline for each packet where $M$ is an integer. Specifically, if a packet is not received correctly by the receiver at the end of the $M^{th}$ slot, it will be discarded from the transmitter queue. During the transfer of each packet, if the packet is successfully decoded at the end of any slot, the receiver feeds the transmitter with an ACK, and the transmitter removes the packet from its queue. Otherwise, the receiver sends a NACK,  and the transmitter, depending on the transmission protocol, either repeats the same encoded packet or sends the next partition of a larger codeword in the next slot. We note that a removal of any data packet from the buffer, either as a result of a successful decoding by the receiver or due to the transmission deadline, is considered to be a packet service from the queue.

All along, the discrete-time channel input-output relation in the $t^{th}$ symbol instant is given as
\begin{equation*}
y(t)=h(t)x(t)+w(t)\quad \text{for }t=0,1,\cdots,
\end{equation*}
where $x(t)$ is the complex channel input and $y(t)$ is the complex channel output. Above, $\{w(t)\}$ forms an independent and identically distributed (i.i.d.) sequence of additive zero-mean circularly symmetric, complex Gaussian random noise variables with a variance $\mathbb{E}\{|w(t)|^2\}=\sigma_w^2$. Furthermore, $h(t)$ denotes the fading coefficient between the transmitter and the receiver with an arbitrary marginal distribution and a finite variance $\mathbb{E}\{|h(t)|^2\}=\mathbb{E}\{z(t)\}=\sigma_h^2<\infty$. Note that, here and throughout the paper, $z(t)=|h(t)|^2$ denotes the magnitude-square of the fading coefficients. We further assume that channel side information (CSI) is available at the receiver, i.e., the receiver knows the actual value of $h(t)$, while the transmitter is only aware of channel statistics. Besides, due to a limited power budget, the channel input is subject to the following average power constraint: $\mathbb{E}\{|x(t)|^2\}\leq\bar{P}/B$ where $B$ is the available channel bandwidth. Since we assume that $B$ complex symbols per second are transmitted, the average power of the system is constrained by $\bar{P}$. Finally, we consider a block-fading channel model and assume that the fading coefficient stays constant for a slot duration of $T$ seconds and changes independently from one slot to another, i.e., $h(lTB)=h(lTB+1)=\cdots=h((l+1)TB-1)=h_{l}$ in the $l^{th}$ slot, and $|h_{l}|^{2}=z_{l}$. This channel model has been widely used to represent slowly-varying, flat fading channels \cite[and references therein]{randal}.
\begin{figure*}
\begin{center}
\includegraphics[width=0.70\textwidth]{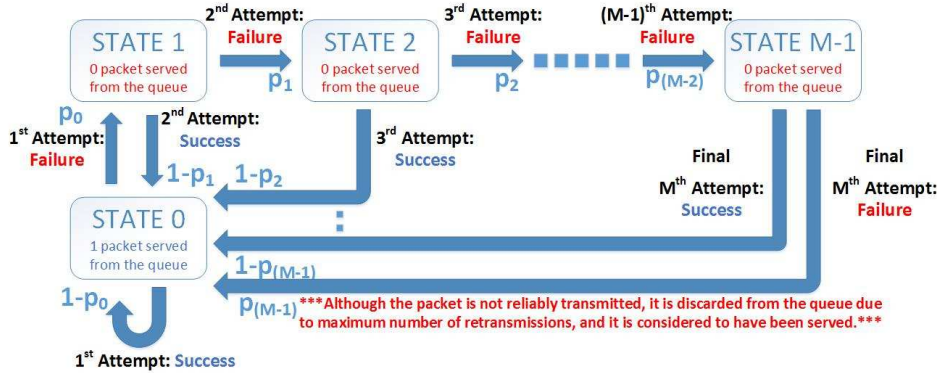}
\caption{State-transition model with $M$ states. The number of bits removed from the transmitter queue is $n$ in State 0, while it is 0 in other states.}\label{res:res_2}
\end{center}
\end{figure*}

In the above model, the instantaneous ergodic channel capacity in the $l^{th}$ slot is given by
\begin{align}
C_{l}=B\log_2\left\{1+\gamma z_{l}\right\}\text{ bits/sec, for }l=0,1,\cdots,\label{ins_capacity}
\end{align}
where the average transmitted signal-to-noise ratio ($\tSNR$) is $\gamma=\frac{\bar{P}}{B\sigma_n^2}$. In (\ref{ins_capacity}), the capacity is achievable when $x(t)$ is zero-mean Gaussian-distributed\footnote{Codewords can be composed of symbols from any modulation technique with the input distribution $Q(x)$. Then, the capacity in (\ref{ins_capacity}) is replaced with the achievable rate expressed as the mutual information between $x$ and $y$, i.e., $I(x;y)$. The physical layer framework of this study and the analysis in the rest of the paper can be easily adopted to any practical linear modulation technique by using $Q(x)$ and $I(x;y)$.}, i.e., $x(t)\sim\mathcal{CN}(0,\bar{P}/B)$ \cite[Ch. 9.1]{book_information_theory}. Since a block-fading channel is considered, we denote the channel fading power and the instantaneous channel capacity in the $l^{th}$ slot by $z_{l}$ and $C_{l}$, respectively. Additionally, we remark that if the transmitter would know the instantaneous channel fading coefficient $h_{l}$, it could send the data with a rate equal to the instantaneous channel capacity. Since in practice $h_{l}$ is generally not available at the transmitter, we consider a more realistic scenario, where a wireless link is equipped with an HARQ protocol, which can be modeled as a discrete-time finite state Markov chain.

\subsection{HARQ}\label{hybrid_arq}
In general, we assume that before being delivered into the channel, a packet of $n$ bits is encoded and modulated into a codeword of length either $TB$ or $MTB$ complex symbols. We can easily deduce that since one fading slot is $T$ seconds, and the available bandwidth is $B$ Hz, a total of $TB$ complex symbols can be transmitted in one slot. Therefore, a codeword of $TB$ complex symbols is assumed to be repeatedly transmitted in each slot until either it is decoded by the receiver or the transmission deadline is reached. HARQ-T1 and HARQ-CC are two of these protocols that perform iterative transmissions of data packets. On the other hand, a codeword of $MTB$ complex symbols is abstracted into $M$ sub-codewords that are transmitted consecutively in each slot again until either it is decoded or the transmission deadline is reached, e.g., HARQ-IR.

Particularly, the first partition (codeword or sub-codeword of length $TB$ complex symbols) is sent to the receiver in the first attempt\footnote{In order to differentiate any slot from the slot in which a packet is transmitted for the first time, we use \emph{attempt} rather than \emph{slot}, unless otherwise needed for clarity.}. If a successful decoding is performed, the receiver sends an ACK to the transmitter, and then the packet is removed from the queue. On the other hand, if a decoding failure occurs, the receiver sends a NACK to the transmitter and requests the second partition. Then, the transmitter sends the second partition in the second attempt. Equivalently, following a successful decoding in any attempt, the packet is removed from the queue; otherwise with a decoding failure, the corresponding next partition is transmitted in the next attempt. However, at the end of the $M^{th}$ attempt, the packet is removed from the queue without waiting for any ACK or NACK since the transmission deadline for a packet is reached. We finally note that due to a decoding failure at the end of the $M^{th}$ attempt, if a packet is removed from the queue and not transmitted again, we deem it as a lost packet at the receiver in order to differentiate it from the other packet removals. So, the ratio of the lost packets to the total number of packets served by the transmitter is a very critical QoS measure since it entails the average reliable transmission rate to the receiver.

\subsection{State Transition Model}\label{state-transition-model}
For an analytical presentation, we model the queue activity at the end of each slot as a discrete-time Markov process. As seen in Fig. \ref{res:res_2}, following a decoding success at the end of the $m^{th}$ attempt, the system enters State 0 with probability 1-$p_{m-1}$, whereas if a decoding failure occurs the system enters State $m$ with probability $p_{m-1}$ where $m\in\{1,\cdots,M-1\}$. On the other hand, regardless of the decoding result at the end of $M^{th}$ attempt, the system goes to State 0 with probability 1 since the transmitted packet is removed from the queue due to the transmission deadline. We note that a packet transmitted in the $M^{th}$ attempt is decoded either correctly with probability 1-$p_{M-1}$ or incorrectly with probability $p_{M-1}$. Accordingly, we have the following state transition matrix $P$:
\begin{equation}\label{PP}
P=\begin{pmatrix}
    1-p_0 & 1-p_1 & 1-p_2 & \cdots & 1-p_{M-2} & 1\\
    p_0 & 0 & 0 & \cdots & 0 & 0\\
    0 & p_1 & 0 & \cdots & 0 & 0\\
    0 & 0 & p_2 & \cdots & 0 & 0\\
    \vdots &\vdots & \vdots& \ddots&\vdots&\vdots\\
    0 & 0 & 0 & \cdots & p_{M-2} & 0
  \end{pmatrix}.\nonumber
\end{equation}

Now, noting that the analysis in the paper is for a general HARQ protocol, we consider the following three different HARQ protocols in a Rayleigh\footnote{The Rayleigh fading channel assumption fits well with HARQ schemes in practical mobile wireless systems \cite{chaitanya}. However, other channel models can be adopted easily into this study.} channel fading environment for a profound examination.

\subsubsection{Type-I HARQ (HARQ-T1)}\label{Protocol_1} In this protocol, the transmitter encodes and modulates its data packets, of $n$ bits, into one codeword by using a channel code of a codebook $\mathcal{C}\in\mathbb{C}^{TB}$ of length $TB$ over the complex numbers. The codeword is transmitted over the channel in one slot. We assume that if the transmission rate is lower than or equal to the instantaneous channel capacity in (\ref{ins_capacity}), i.e., $\frac{n}{T}\leq C_{l}$, the receiver is able to decode the codeword reliably. On the other hand, if the transmission rate is greater than the capacity, i.e., $\frac{n}{T}>C_{l}$, the receiver cannot recover the packet correctly. Featuring a receiver that does not perform accumulation of data, i.e., that uses only the current received signal in any slot in order to retrieve the packet, the decoding failure probability in any slot is given by
\begin{align*}
p=\Pr\left\{\frac{n}{T}>C_{l}\right\}=\Pr\{\kappa>z_{l}\}=1-e^{-\kappa/\sigma_{h}^{2}}
\end{align*}
where the probability density function of exponentially-distributed $z_{l}$ (i.e., Rayleigh distributed $h_{l}$) is $f_{z_{l}}(z_{l})=e^{-z_{l}/\sigma_{h}^{2}}$ and $\kappa=(2^{\frac{n}{TB}}-1)/\gamma$. Since any state transition does not depend on the past transitions, we have $p=p_0=\cdots=p_{M-1}$.
\subsubsection{Type-II HARQ-CC}
In this protocol, we again assume that the transmitter encodes and modulates its data packets, of $n$ bits each, by using a channel code of a codebook $\mathbb{C}\in\mathcal{C}^{TB}$ of length $TB$ over the complex numbers. However, different than the receiver structure in Section \ref{Protocol_1}, we devise a receiver that makes use of the received signals in the earlier slots by implementing maximum ratio combining. Hence, the average accumulated mutual information at the receiver at the end of the $m^{th}$ slot for $1\leq m\leq M$ is expressed as \cite{chaitanya}
\begin{equation*}
C_{l}^{l+m-1}=\frac{B}{m}\log_{2}\left\{1+\gamma\sum_{i=0}^{m-1}z_{l+i}\right\}
\quad\text{bits/sec}.
\end{equation*}
Above, while $C_{l}^{l}$ is the channel capacity in the first slot, $C_{l}^{l+k-1}$ can be considered as the average channel capacity in the first $k$ slots. Noting that the receiver can decode a packet at the end of $m^{th}$ slot when $\frac{n}{mT}\leq C_{l}^{l+m-1}$, we have the following state transition probabilities:
\begin{align*}
p_{m-1}&=\Pr\left\{\frac{n}{mT}>C_{l}^{l+m-1}|\frac{n}{(m-1)T}>C_{l}^{l+m-2}\right\}\nonumber\\
&=\frac{\Pr\left\{\frac{n}{mT}>C_{l}^{l+m-1}\right\}}{\Pr\left\{\frac{n}{(m-1)T}>C_{l}^{l+m-2}\right\}}\\
&=\frac{\Pr\left\{\kappa>\sum_{i=0}^{m-1}z_{l+i}\right\}}{\Pr\left\{\kappa>\sum_{i=0}^{m-2}z_{l+i}\right\}}=\frac{\gamma\left(m,\kappa/\sigma_{h}^{2}\right)}{(m-1)\gamma\left(m-1,\kappa/\sigma_{h}^{2}\right)}
\end{align*}
where $m\in\{2,\cdots,M\}$, and $p_{0}=1-e^{-\kappa/\sigma_{h}^{2}}$. Above, $\gamma(a,b)$ is the lower incomplete gamma function where $\gamma(a,b)=\int_{0}^{b}t^{a-1}e^{-t}dt$, and $\kappa$ is as defined in Section \ref{Protocol_1}. Note that the sum of exponentially-distributed random variables is gamma-distributed.
\subsubsection{Type-II HARQ-IR}
In this protocol, different than the other two protocols, we assume that the transmitter encodes and modulates its data packets, of $n$ bits each, by using a channel code of a codebook $\mathbb{C}\in\mathcal{C}^{MTB}$ of length $MTB$ over the complex numbers. Then, the transmitter divides the codewords into $M$ partitions of the same length with $TB$ complex symbols. In each slot, one partition is sent to the receiver, and the receiver decodes the message using the current partition combined with the previously received partitions related to the encoded data packet. Hence, following the result in \cite{caire}, we can see that the receiver can decode the transmitted data packet at the end of the $m^{th}$ slot, if $\frac{n}{T}\leq \sum_{i=0}^{m-1}B\log_{2}\left(1+\gamma z_{l+i}\right)$. Subsequently, we can express the state transition probabilities for $m\in\{2,\cdots,M\}$ as follows:
\begingroup
\allowdisplaybreaks
\begin{align*}
p_{m-1}=\frac{\Pr\left\{\gamma\kappa+1>\prod_{i=0}^{m-1}\left(1+\gamma z_{l+i}\right)\right\}}{\Pr\left\{\gamma\kappa+1>\prod_{i=0}^{m-2}\left(1+\gamma z_{l+i}\right)\right\}}
\end{align*}
\endgroup
where $p_{0}=1-e^{-\kappa/\sigma_{h}^{2}}$, and $\kappa$ is as defined in Section \ref{Protocol_1}.
\section{System Performance}\label{system_throughput}
In this section, we concentrate on the performance measures of the HARQ systems with the above defined state-transition model. We initially obtain the packet-loss probability at the receiver, and then determine the effective capacity which provides the maximum sustainable arrival rate at the buffer under certain QoS constraints. We finally provide the non-asymptotic performance measures in the form of backlog and delay bounds.  
\subsection{Packet-Loss Probability}
Let $\mathbf{\pi}=[\pi_{0},\cdots,\pi_{M-1}]^{T}$ be the steady-state probability vector of the above state transition model in Fig. \ref{res:res_2} where $\sum_{j=0}^{M-1}\pi_{j}=1$ and $\mathbf{\pi}=P\mathbf{\pi}$. Thus, we can obtain the steady-state probabilities as follows:
\begin{equation*}
\pi_i=\pi_0\prod_{j=0}^{i-1}p_{j}\text{ and }\pi_{0}=\frac{1}{1+\sum_{i=1}^{M-1}\prod_{j=0}^{i-1}p_{j}}
\end{equation*}
where $i\in\{1,\cdots,M-1\}$. Recalling that at the end of the $M^{th}$ slot, data packets are removed from the queue due to a decoding failure with probability $p_{M-1}$ and due to a decoding success with probability $1-p_{M-1}$, we can easily express the packet-loss probability, i.e., the ratio of the lost packets to the total number of packets that are served by the transmitter, with
\begin{align*}
p_{\text{lost}}&=\frac{p_{M-1}\pi_{M-1}}{\pi_{0}}=p_{M-1}\prod_{j=0}^{M-2}p_{j}=\prod_{j=0}^{M-1}p_{j}.
\end{align*}
Note that while $\pi_0$ is the steady-state probability of removing a packet from the queue in one slot, $p_{\text{lost}}$ is the probability of a packet not reaching the receiver.
\begin{figure}
\begin{center}
\includegraphics[width=\figsize\textwidth]{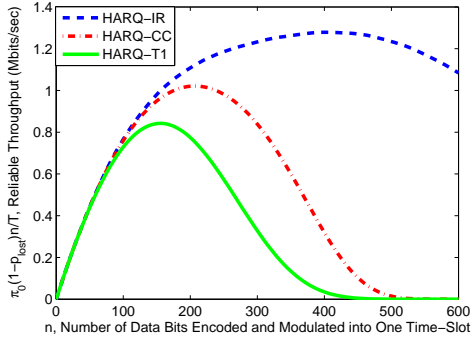}
\caption{Reliable throughput, $\frac{\pi_{0}(1-p_{\text{lost}})n}{T}$ bits/sec., v.s. number of bits encoded and modulated into $TB$ symbols in one slot, $n$, when $\gamma=5$ dB.}\label{fig:fig_2_add}
\end{center}
\end{figure}

\begin{figure}
\begin{center}
\includegraphics[width=\figsize\textwidth]{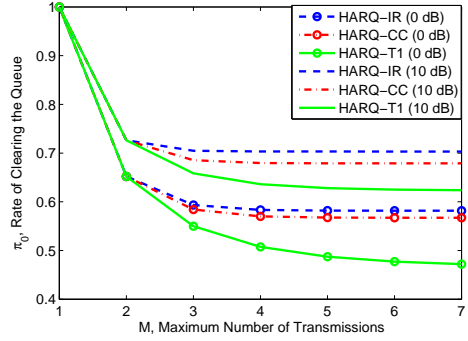}
\caption{Steady state probability of State 0, $\pi_0$, as a function of transmissions deadline, $M$, for three different protocols when $\gamma=0$ and $10$ dB.}\label{fig:fig_1}
\end{center}
\end{figure}

\begin{figure}
\begin{center}
\includegraphics[width=\figsize\textwidth]{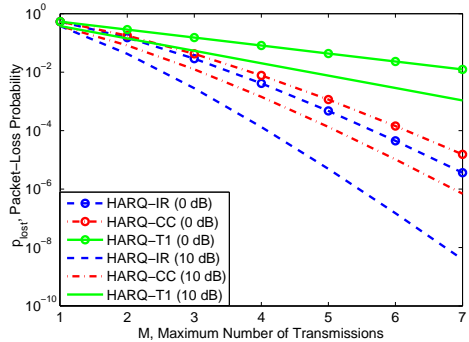}
\caption{Packet-loss probability, $p_{\text{lost}}$, as a function of transmissions deadline, $M$, for three different protocols when $\gamma=0$ and $10$ dB.}\label{fig:fig_2}
\end{center}
\end{figure}

We employ the aforementioned HARQ protocols, and plot the average reliable throughput that arrives at the receiver, i.e., $\frac{\pi_{0}(1-p_{\text{lost}})n}{T}$ bits/sec, as a function of $n$, i.e., the number of bits that are encoded and modulated into $TB$ symbols in one slot, when $\gamma=5$ dB in Fig. \ref{fig:fig_2_add}. We consider the following settings: The average channel fading power is $\sigma_{h}^{2}=1$, the slot duration is $T=100$ $\mu$sec, and the transmission bandwidth is $B=1$ MHz. We observe that while HARQ-IR significantly outperforms the other two protocols, HARQ-T1 is the least achieving protocol for all $n$. Fixing a constant number of bits served in all protocols\footnote{Unless stated otherwise, here and throughout the paper, we consider the given transmission settings, and choose $n$ that maximizes the reliable throughput at the receiver when HARQ-T1 is employed.}, we further compare the queue clearing performances of all the protocols, $\pi_0$, as a function of the transmission deadline, $M$, when $\gamma=0$ and 10 dB in Fig. \ref{fig:fig_1}. We can see for any $\gamma$ that the probability $\pi_0$ decreases generally with increasing $M$, then it converges after certain $M$ in all HARQ protocols. All along, HARQ-IR converges very quickly, while HARQ-CC performs very close to HARQ-IR. On the other hand, HARQ-T1 accomplishes considerably less in general compared to the Type-II HARQ protocols for all $\gamma$ values. Finally, we plot the packet-loss probability as a function of $M$ in Fig. \ref{fig:fig_2}. Unlike $\pi_0$, $p_{\text{lost}}$ is continuously decreasing with increasing $M$. HARQ-IR has a significantly lower packet-loss probability, and the performance gap between HARQ-IR and the others increases dramatically with again increasing $M$.

\subsection{Effective Capacity}
Even though it seems to be advantageous to increase $M$ in order to decrease $p_{\text{lost}}$, we can promptly infer that the average delay of a packet in the buffer will increase with increasing $M$. Therefore, we propose the effective capacity that characterizes the asymptotic decay rate of buffer occupancy. It identifies the maximum constant arrival rate that a given service process can support in order to guarantee a desired statistical QoS specified with the QoS exponent $\theta$ \cite{wu_negi}. Defining $Q(t)$ as the stationary queue length at time $t$, and $\theta$ as the decay rate of the tail distribution of the queue length $Q(t)$, we can express the following: $\lim_{q\to\infty}\frac{\log{\Pr(Q(t)\geq q)}}{q} = -\theta$. Therefore, we have the following approximation for larger $q$: $\Pr(Q(t)\geq q)\approx e^{-\theta q}$, which means that larger $\theta$ refers to strict buffer constraints, and smaller $\theta$ implies looser constraints. Furthermore, it is shown in \cite{liu_chamberland} that $\Pr(D(t)\geq d)\leq c\sqrt{\Pr(Q(t)\geq q)}$ for constant arrival rates, where $D(t)$ denotes the steady-state delay experienced in the buffer, and $c$ is a positive constant. In the above formulation, we have $q=ad$. Therefore, effective capacity provides us with the maximum arrival rate when the system is subject to the statistical queue length or delay constraints in the forms of $\Pr(Q(t)\geq q)\leq e^{-\theta q}$ or $\Pr(D(t)\geq d)\leq ce^{-\theta ad/2}$, respectively. For a given QoS exponent $\theta>0$, effective capacity, $\rho_{S}(\theta)$, is given by
\begin{equation}\label{for_slack_term}
\rho_{S}(\theta)=-\frac{\Lambda(-\theta)}{\theta}=-\lim_{t\to\infty}\frac{1}{\theta t}\log E\{e^{-\theta S(0,t)}\}
\end{equation}
where $S(\tau,t)=\sum_{k=\tau+1}^{t}r(k)$ is the time-accumulated service process, and $r(k)$ for $k=1,2,...$ is the discrete-time, stationary and ergodic stochastic service process. We note that $\Lambda(\theta)$ is the asymptotic log-moment generating function of $S(0,t)$, and is given by $\Lambda(\theta)=\lim_{t\to\infty}\frac{1}{t}\log E\left\{e^{\theta S(0,t)}\right\}$. Henceforth in the next result, we provide the effective capacity for any given HARQ protocol.
\begin{theo}\label{theo:effective_capacity}
For the HARQ system with the state-transition model given in Section \ref{state-transition-model}, the effective capacity for a given QoS exponent $\theta$ is given by
\begin{equation}\label{effective_capacity_rho}
\rho_{S}(\theta)=-\frac{1}{\theta T}\log\left(p_{0}e^{-n\theta}y^{\star}\right)\text{ bits/sec}
\end{equation}
where $y^{\star}$ is the only unique real positive root of $f(y)$ where
%\begingroup
%\allowdisplaybreaks
\begin{align}\label{f_yy}
\begin{split}
f(y)=&y^{M}-\frac{1-p_{0}}{p_{0}}y^{M-1}-\frac{p_{M-2}\cdots p_{1}}{p_{0}^{M-1}e^{-(M-1)n\theta}}\\-&\sum_{i=1}^{M-2}\frac{(1-p_i)p_{i-1}\cdots p_{1}}{p_0^{i}e^{-in\theta}}y^{M-1-i}.
\end{split}
\end{align}
%\endgroup
\end{theo}
\emph{Proof:} See Appendix \ref{app:effective_capacity}.

The real positive root of $f(y)$ can be found by using numerical techniques. For instance, bisection method can be efficiently used to find the solution since it has only one real positive root.

\begin{figure}
\begin{center}
\includegraphics[width=\figsize\textwidth]{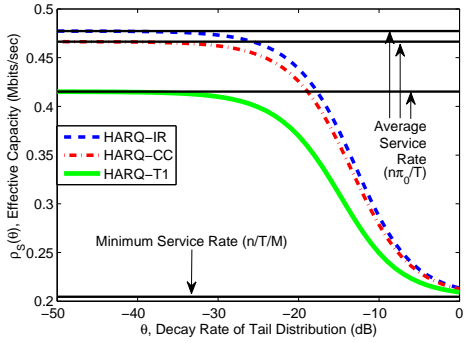}
\caption{Effective capacity, $\rho_{S}(\theta)$, as a function of decay rate of the tail distribution, $\theta$, for three different protocols when $\gamma=0$ dB.}\label{fig:fig_3}
\end{center}
\end{figure}
\begin{figure}
\begin{center}
\includegraphics[width=\figsize\textwidth]{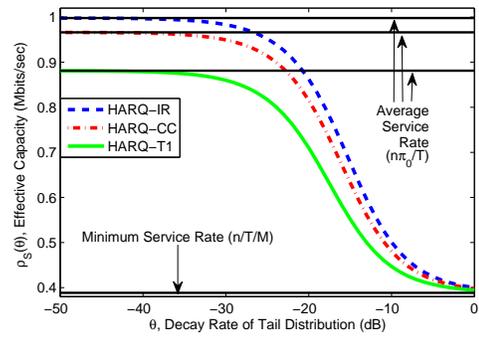}
\caption{Effective capacity, $\rho_{S}(\theta)$, as a function of decay rate of the tail distribution, $\theta$, for three different protocols when $\gamma=5$ dB.}\label{fig:fig_4}
\end{center}
\end{figure}

Employing, again, the above HARQ protocols, we plot the effective capacity, $\rho_{S}(\theta)$, as a function of the decay rate, $\theta$, when $\gamma=0$ dB and $\gamma=5$ dB in Fig. \ref{fig:fig_3} and Fig. \ref{fig:fig_4}, respectively. Above, we set the transmission deadline to $M=4$. In both figures, the superiority of HARQ-IR is clearly seen, and HARQ-CC performs close to HARQ-IR. We further see that while the effective capacity of each protocol goes to the average service rate $\frac{n\pi_{0}}{T}$ of the same protocol with decreasing $\theta$, all of them go to the same minimum service rate $\frac{n}{TM}$ with increasing $\theta$.

\subsection{Non-asymptotic Performance Bounds}
So far, we have considered the effective capacity that deals with asymptotic characteristics. However, non-asymptotic performance bounds regarding the statistical characterizations of backlog and delay are also of interest for system designers. For a non-asymptotic analysis we use the framework of the stochastic network calculus\cite{Chang,ciucu,jiang,fidler_rizk}. Following the model in \cite[Definition 7.2.1]{Chang}, we define a statistical affine bound for the above channel model for any $\theta$ as follows:
\begin{equation}\label{affine_bound}
E\left\{e^{-\theta S(\tau,t)}\right\}\leq e^{-\theta[\rho_{S}(\theta)(t-\tau)-\sigma_{S}(\theta)]}
\end{equation}
where $\rho_{S}(\theta)$ is the effective capacity and $\sigma_{S}(\theta)$ is a slack term that defines an initial service delay. Although the expression in (\ref{affine_bound}) seems to be an upper bound, due to $-\theta$, in fact (\ref{affine_bound}) is a lower bound on the expected amount of service. Now, using Chernoff's lower bound $\Pr\left\{X\leq x\right\}\leq e^{\theta x}E\left\{e^{-\theta X}\right\}$ for $\theta\geq0$, the so-called exponentially bounded fluctuation model described in \cite{klee} with parameters $\rho_{S}(\theta)>0$, $b\geq0$, and $\Pr\left\{S(\tau,t)<\rho_{S}(\theta)(t-\tau)-b\right\}\leq \varepsilon(b)$ follows, where $\varepsilon(b) = e^{\theta \sigma_S(\theta)}e^{-\theta b}$ specifies an exponentially decaying deficit profile of the service. A sample path guarantee follows by the use of the union bound as:
\begin{equation*}
\Pr\left\{\exists\tau\in[0,t]:S(\tau,t)<\rho_{S}'(\theta)(t-\tau)-b\right\}\leq\varepsilon'(b)
\end{equation*}
where
\begin{equation}\label{error_bb}
\varepsilon'(b)=\frac{e^{\theta\sigma_{S}(\theta)}}{1-e^{-\theta\delta}}e^{-\theta b}
\end{equation}
and $\rho_{S}'(\theta)=\rho_{S}(\theta)-\delta$ with free parameter $0<\delta\leq \rho_{S}(\theta)-a$ for an arrival rate $a$. For a detailed derivation, we refer the interested reader to \cite{fidler_rizk}.
\begin{figure}
\begin{center}
\includegraphics[width=\figsize\textwidth]{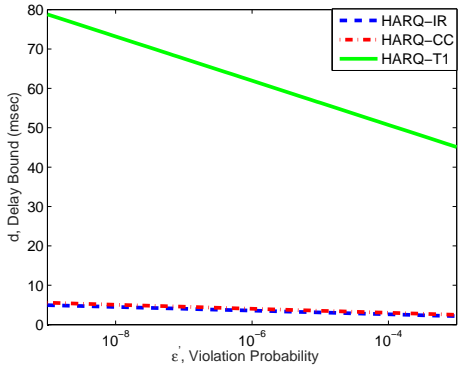}
\caption{Delay bound, $d$, as a function of the violation probability, $\varepsilon'$, for three different protocols when $\gamma=0$ and $a=0.41$ Mbits/sec with $n=82$.}\label{fig:fig_5}
\end{center}
\end{figure}
\begin{figure}
\begin{center}
\includegraphics[width=\figsize\textwidth]{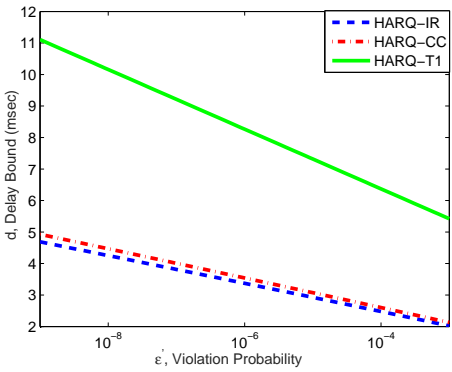}
\caption{Delay bound, $d$, as a function of the violation probability, $\varepsilon'$, for three different protocols when $\gamma=5$ and $a=0.81$ Mbits/sec and $n=155$.}\label{fig:fig_6}
\end{center}
\end{figure}
All in all, perceive that the backlog $Q(t) = \max_{\tau \in [0,t]} \{a(t-\tau)-S(\tau,t)\}$ has a statistical bound $q = \max_{\tau\in[0,t]}\{a(t-\tau)-[\rho_{S}'(\theta)(t-\tau)-b]_{+}\}$ that may fail with probability $\Pr\{Q(t) > q\} \le \varepsilon'(b)$, where $[x]_{+}=0$ if $x<0$, and $[x]_{+}=x$ otherwise accounts for the fact that $S(\tau,t) \geq 0$, generally. If $a\leq\rho_{S}'(\theta)$ for stability,
\begin{equation}\label{add_ext}
q=a\frac{b}{\rho_{S}(\theta)-\delta}
\end{equation}
follows for all $t$. Accordingly, the delay bound $\Pr\{D(t)>d\}\leq\varepsilon'(b)$ can be expressed with $d=q/a$. In (\ref{add_ext}), $b/(\rho_{S}(\theta)-\delta)$ provides us with the initial latency caused by the variability of the service. Finally, by inversion of (\ref{error_bb}), $b$ can be easily obtained for any given $\varepsilon'$ by
\begin{equation}\label{f_b}
b=\sigma_{S}(\theta)-\frac{1}{\theta}\left(\log(\varepsilon')+\log\left(1-e^{-\theta\delta}\right)\right).
\end{equation}

As for the existence of the slack term $\sigma_{S}(\theta)$ in (\ref{f_b}), considering the result \cite[7.2.6.ii]{Chang}, we provide the following lemma:
\begin{lemm}\label{lemm:slack_term}
If $S(\tau,t)$ has an envelope rate $\rho_{S}(\theta)<\infty$, for every $\epsilon>0$, there exists $\sigma_{S}(\theta)<\infty$ such that $S(\tau,t)$ is $(\sigma_{S}(\theta),\rho_{S}(\theta)-\epsilon)$-upper constrained (\ref{affine_bound}).
\end{lemm}
\emph{Proof:} See Appendix \ref{app:slack_term}.

\begin{figure}
\begin{center}
\includegraphics[width=\figsize\textwidth]{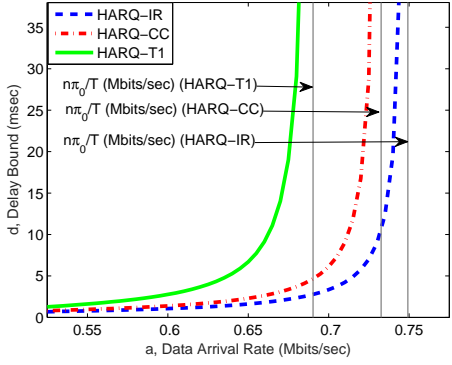}
\caption{Delay bound, $d$, as a function of data arrival rate, $a$, when $\gamma=5$ dB.}\label{fig:fig_add_1}
\end{center}
\end{figure}

\begin{figure}
\begin{center}
\includegraphics[width=\figsize\textwidth]{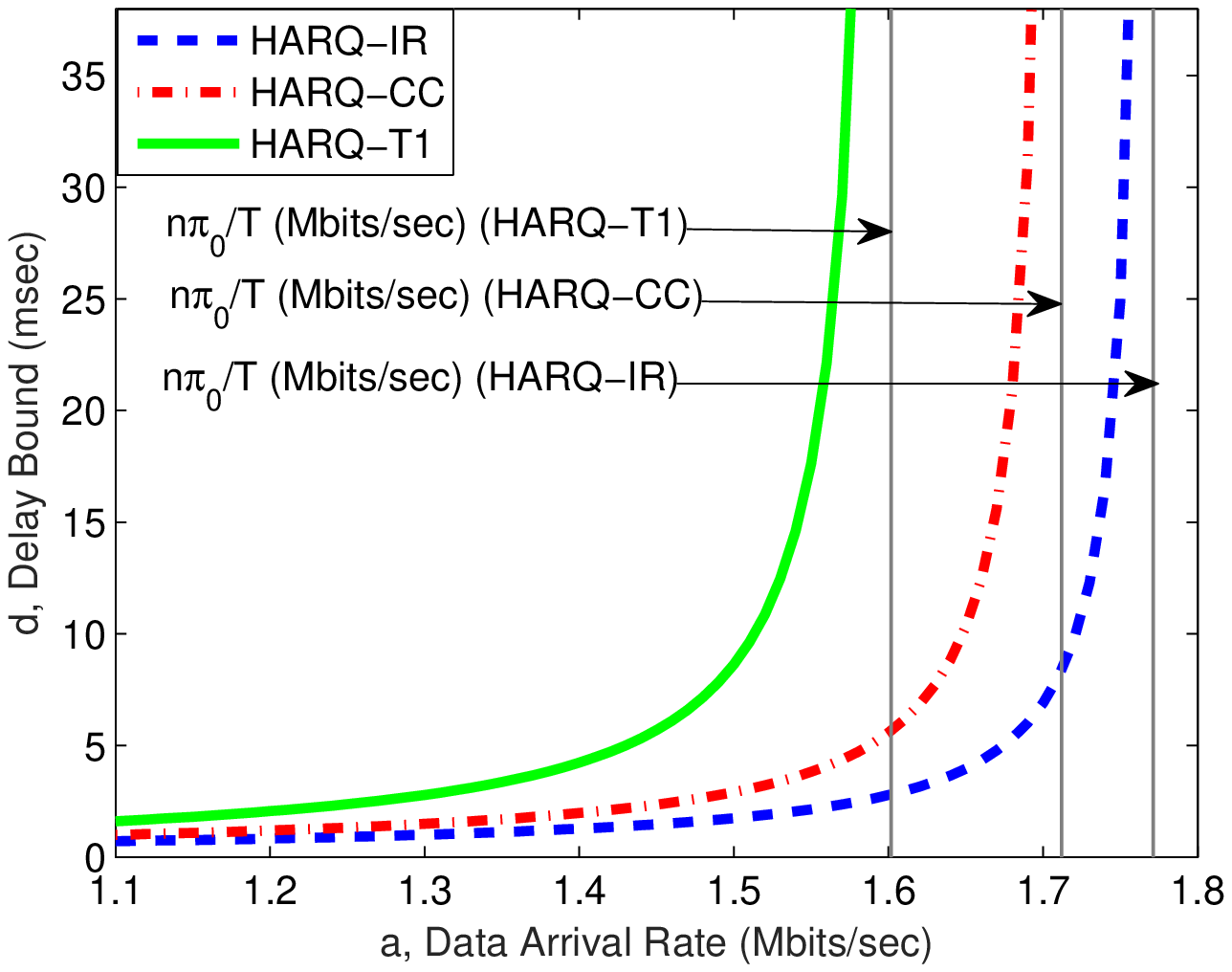}
\caption{Delay bound, $d$, as a function of data arrival rate, $a$, when $\gamma=10$ dB.}\label{fig:fig_add_2}
\end{center}
\end{figure}

As seen in Fig. \ref{fig:fig_5} and Fig. \ref{fig:fig_6}, we plot the delay bound threshold, $d$, as a function of the violation probability, $\varepsilon'$, when $\gamma=0$ dB and $\gamma=5$ dB when the arrival rates are $a=0.41$ Mbits/sec and $a=0.81$ Mbits/sec while $n=82$ bits/slot and $n=155$ bits/slot are encoded and modulated into 100 symbols, respectively. We observe that HARQ-T1 has a very poor performance when compared with the other Type-II HARQ protocols. For instance, $d$ is around 78 msec when HARQ-T1 is utilized, whereas it is around 5 msec when the Type-II HARQ protocols are engaged for $\varepsilon=10^{-9}$ and $\gamma=0$ dB. Similarly, when $\gamma=5$ dB, $d$ is reduced from around 11 msec to around 4 msec. We remark that the arrival rates are different and are close to the average service rates in both plots. We again observe that HARQ-CC and HARQ-IR achieve delay bounds very close to each other.
\begin{figure}
\begin{center}
\includegraphics[width=\figsize\textwidth]{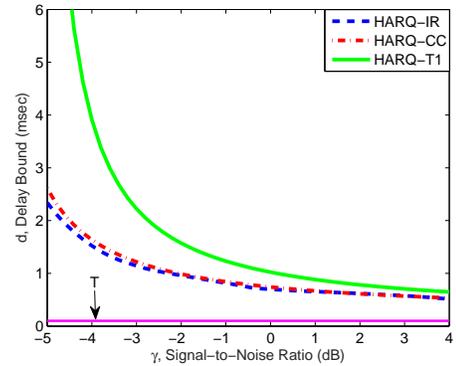}
\caption{Delay bound, $d$, as a function of signal-to-noise ratio, $\gamma$, when $M=4$ and $a=0.16$ Mbits/sec.}\label{fig:fig_add_3}
\end{center}
\end{figure}
\begin{figure}
\begin{center}
\includegraphics[width=\figsize\textwidth]{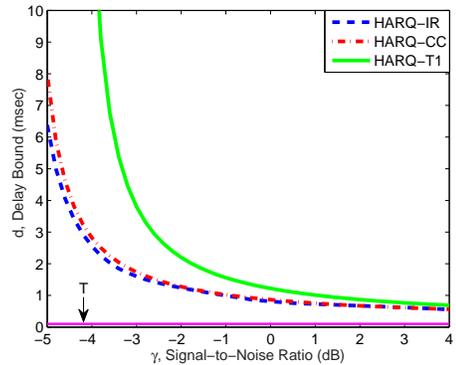}
\caption{Delay bound, $d$, as a function of signal-to-noise ratio, $\gamma$, when $M=3$ and $a=0.18$ Mbits/sec.}\label{fig:fig_add_4}
\end{center}
\end{figure}
Furthermore, setting a fixed arrival rate $a=0.16$ Mbits/sec and $a=0.18$ Mbits/sec, we plot the delay bound threshold, $d$, as a function of $\tSNR$, $\gamma$, in Fig. \ref{fig:fig_add_3} and Fig. \ref{fig:fig_add_4} when $M=4$ and $3$, respectively, while $\varepsilon'=10^{-6}$ and $n=36$ bits/slot. As expected, the delay bound decreases with increasing $\gamma$. However, the decrease in $d$ converges in all HARQ protocols. This is due to the fixed transmission slot length, $T$. It is worth mentioning that when the delay is the main concern, expending transmission energy above some value is useless, and that when there is enough energy provided for data transmission, employing a simplified technology is more profitable for system designers. As for the delay performance regarding the variations in data arrival rates at the buffer, we plot $d$ as a function of $a$ in Fig. \ref{fig:fig_add_1} and Fig. \ref{fig:fig_add_2} when $\gamma=5$ dB and $10$ dB while $n=155$ bits/slot and $252$ bits/slot, respectively. We again note $\varepsilon'=10^{-6}$. In all protocols and cases, when the arrival rate is greater than the average service rate, the delay threshold goes to infinity since the system is not stable any more. The performance of HARQ-T1 is noticeably degraded with increasing $a$ when compared to other HARQ protocols. However, a system furnished with HARQ-T1 is more appreciated when arrival rates are lower.

\section{Conclusion}\label{conclusion}
In this paper, we have considered HARQ systems under QoS constraints such as asymptotic and non-asymptotic delay and backlog bounds. We have addressed a practical setting in which data packet transmission under a transmission deadline is performed. We have initially established a state-transition model to define the queueing characteristics of the system, then we have identified the steady-state probability of clearing the queue at the transmitter and the packet-loss probability at the receiver. We have further derived the effective capacity that features the arrival rate at the transmitter queue under QoS constraints. We have then furnished the analysis with the non-asymptotic delay and backlog bounds with respect to a given violation probability. We have finally presented the numerical results comparing three different HARQ protocols.

\appendix

\subsection{Proof of Theorem \ref{theo:effective_capacity}}\label{app:effective_capacity}
In \cite[Chap. 7, Example 7.2.7]{Chang}, it is shown for Markov modulated processes that
\begin{equation*}
\frac{\Lambda(\theta)}{\theta}=\frac{1}{\theta}\log\ssp\{P\phi(\theta)\}=\frac{1}{\theta}\log\ssp\{\Upsilon\}
\end{equation*}
where $\ssp\{\Upsilon\}$ is the spectral radius of the matrix $\Upsilon$, $P$ is the transition matrix of the underlying Markov process, and $\phi(\theta)=\diag\{\phi_{0}(\theta),...,\phi_{M-1}(\theta)\}$ is a diagonal matrix, components of which are the moment generating functions of the processes in $M$ states. The rates (number of bits leaving the queue) supported by the above channel model with the state transition model described in the previous section can be seen as a Markov modulated process and hence the setup considered in \cite{Chang} can be applied immediately into our setting. We have $n$ bits served from the queue in State 0 while we have 0 bit served in other states. Therefore, we have $\phi(\theta)=\diag\{e^{-n\theta},1,...,1\}$. Then, we have
\begin{align}\label{transition_matrix_phi}
\Upsilon=&\begin{pmatrix}
    (1-p_0)e^{-n\theta} & 1-p_1 & \cdots & 1-p_{M-2} & 1\\
    p_0e^{-n\theta} & 0 & \cdots & 0 & 0\\
    0 & p_1 & \cdots & 0 & 0\\
    \vdots &\vdots & \ddots&\vdots&\vdots\\
    0 & 0 & \cdots & p_{M-2} & 0
  \end{pmatrix}.
\end{align}
Note that $\Upsilon$ in (\ref{transition_matrix_phi}) is a non-negative matrix, i.e., $\Upsilon\geq0$, with each element greater or equal to zero, i.e., $\upsilon_{ij}\geq0$, where $\Upsilon=[\upsilon_{ij}]$. \cite[Chap. 8, Corollary 8.3.3]{Horn} states that if $\Upsilon\geq0$, $\mathbf{x}=[x_{0},\cdots,x_{M-1}]\geq0$, and $\mathbf{x}\neq0$ then
\begingroup
\allowdisplaybreaks
\begin{align}
\ssp(\Upsilon)=&\max_{\substack{\mathbf{x}\geq0\\\mathbf{x}\neq0}}\min_{\substack{0\leq i\leq M-1}}\frac{1}{x_{i}}\sum_{j=0}^{M-1}\upsilon_{ij}x_{j}\nonumber\\
=\max_{\substack{\mathbf{x}\geq0\\\mathbf{x}\neq0}}\Bigg[\min&\Bigg\{(1-p_0)e^{-n\theta}+\frac{x_{M-1}+\sum_{i=1}^{M-2}(1-p_i)x_i}{x_{0}},\nonumber\\
p_0e^{-n\theta}\frac{x_0}{x_1},&p_1\frac{x_1}{x_2},\cdots,p_{M-2}\frac{x_{M-2}}{x_{M-1}}\Bigg\}\Bigg].\label{rho_2}
\end{align}
\endgroup
Now, let us assume that the maximum of the minimum in (\ref{rho_2}) is obtained whenever
\begin{align}\label{first_ssp}
\begin{split}
p_0e^{-n\theta}&\frac{x_0}{x_1}=p_1\frac{x_1}{x_2}=\cdots=p_{M-2}\frac{x_{M-2}}{x_{M-1}}\\
&=(1-p_0)e^{-n\theta}+\frac{x_{M-1}+\sum_{i=1}^{M-2}(1-p_i)x_i}{x_{0}}.
\end{split}
\end{align}
Note that whenever $x_i$ for $i=0,\cdots,M-1$ changes, the minimum in (\ref{rho_2}) will decrease, but it will not be the maximum of the minimum values. Therefore, $\ssp(\Upsilon)$ will be obtained whenever (\ref{first_ssp}) is provided. Holding the equality in (\ref{first_ssp}), we have
\begin{equation}\label{relation}
x_{i}=x_{0}\frac{p_{i-1}p_{i-2}\cdots p_1}{p_{0}^{i-1}e^{(1-i)n\theta}}\left(\frac{x_{1}}{x_{0}}\right)^{i}\quad \text{for i=2$\cdots$M-1}.
\end{equation}
Then, defining $y=\frac{x_0}{x_1}$, and inserting $y$ and (\ref{relation}) into the equality in (\ref{first_ssp}), we have $\ssp(\Upsilon)=p_0e^{-n\theta}y$ and
\begin{align}\label{func_y}
\begin{split}
y=&\frac{1-p_0}{p_0}+\sum_{i=1}^{M-2}\frac{(1-p_i)p_{i-1}\cdots p_{1}}{p_0^{i}e^{-in\theta}}\frac{1}{y^{i}}\\&+\frac{p_{M-2}\cdots p_1}{p_{0}^{M-1}e^{-(M-1)n\theta}}\frac{1}{y^{M-1}}.
\end{split}
\end{align}
Reordering (\ref{func_y}), we obtain the expression $f(y)$ in (\ref{f_yy}). In order to analyze the roots of $f(y)$, we invoke the following theorem:
\begin{theo}[Cauchy's Theorem]
Let $g(x)=x^{n}-b_1x^{n-1}-\cdots-b_n$, where all the numbers $b_i$ are non-negative and at least one of them is non-zero. The polynomial $g(x)$ has a unique positive root $x^{*}$ and the absolute values of the other roots do not exceed $x^{*}$ \cite{Prasolov}.
\end{theo}

Note that $f(y)$ has coefficients that are non-negative and at least one of them is non-zero. Therefore, there is only one unique real positive root of $f(y)$, denoted by $y^{\star}$, which gives the spectral radius of $\Upsilon$.

\subsection{Proof of Lemma \ref{lemm:slack_term}}\label{app:slack_term}
We can note from (\ref{for_slack_term}) that for every $\epsilon>0$, there exists $t_{0}<\infty$ such that
\begin{align}
\frac{1}{\theta}\sup_{\tau\geq0}\log Ee^{-\theta S(\tau,\tau+t)}&\leq t(-\rho_{S}(\theta)+\epsilon)\nonumber\\
&\leq t_{0}(-\rho_{S}(\theta)+\epsilon)\label{ek_1}
\end{align}
when $t>t_{0}$. Meanwhile, for all $t\leq t_{0}$, we can write
\begin{align*}
\frac{1}{\theta}\sup_{\tau\geq0}\log Ee^{-\theta S(\tau,\tau+t)}&\leq 0.
\end{align*}
Let $\sigma_{S}(\theta)=t_{0}(\rho_{S}(\theta)-\epsilon)$ such that for $t\leq t_{0}$
\begin{align}\label{upper_constrained_2}
\hspace{-0.4cm}\frac{1}{\theta}\sup_{\tau\geq0}\log Ee^{-\theta S(\tau,\tau+t)}\leq0\leq t(-\rho_{S}(\theta)+\epsilon)+\sigma_{S}(\theta).
\end{align}
From (\ref{ek_1}) and (\ref{upper_constrained_2}), we can easily see that $S(\tau,\tau+t)$ is $(\sigma_{S}(\theta),\rho_{S}(\theta)-\epsilon)$-upper constrained, i.e., for all $0\leq t$
\begin{align*}
\sup_{\tau\geq0}\log Ee^{-\theta S(\tau,\tau+t)}\leq-\theta\left[t(\rho_{S}(\theta)-\epsilon)-\sigma_{S}(\theta)\right].
\end{align*}

\bibliographystyle{IEEEtran}
\bibliography{IEEEabrv,references}

\end{document}